\documentclass[12pt]{article}
\pdfoutput=1
\usepackage[a4paper]{geometry}
\usepackage[labelfont=bf]{caption}
\usepackage{subcaption}

\pdfoutput=1

\usepackage{comment}

\usepackage{jheppub,hyperref,float,array,adjustbox,mathtools,physics, xcolor}
\usepackage{graphicx,latexsym} 
\usepackage{amssymb,amsmath}
\usepackage{amsthm,lscape}
\usepackage{longtable,booktabs,setspace} 
\usepackage{url}
\usepackage{xcolor}

\definecolor{bittersweet}{rgb}{1.0, 0.44, 0.37}
\definecolor{ufogreen}{rgb}{0.19, 0.78, 0.43}
\usepackage[bbgreekl]{mathbbol}

\usepackage{amsmath,amssymb,amsfonts,amsxtra, mathrsfs, makeidx,graphicx,amsthm,epsfig, bm,longtable,float, color,tikz,mathtools,xfrac,footnote,rotating, lscape, makecell, environ,mathtools, empheq, physics,cleveref,tensor,slashed,subfiles,natbib}

\usepackage{bbold}
\DeclareSymbolFontAlphabet{\mathbb}{AMSb}
\DeclareSymbolFontAlphabet{\mathbbl}{bbold}

\NewDocumentCommand{\fluxx}{ O{} O{} m m}{[\rho_{#3}^{#1}\rho_{#4}^{#2}]^{y_S}_{y_N}}

\title{
\begin{center}
S-duality in the Cardy-like limit of the superconformal index
\end{center}
}

\author[a]{Antonio Amariti,}	
\author[a,b]{Andrea Zanetti}

\affiliation[a]{INFN, Sezione di Milano, Via Celoria 16, I-20133 Milano, Italy}
\affiliation[b]{Dipartimento di Fisica, Università degli studi di Milano, Via Celoria 16, I-20133}

\emailAdd{antonio.amariti@mi.infn.it}
\emailAdd{andrea.Zanetti@mi.infn.it}

\abstract{
We evaluate the  superconformal index of 
4d $\mathcal{N}=4$ SYM with gauge algebra $so(2N_c+1)$
in the Cardy-like limit.
We then study the relation with the results  obtained for the S-dual $usp(2N_c)$,
discussing the fate of S-duality in different regions of charges. 
We find that S-duality is preserved thanks to a non-trivial integral identity that relates 
the three sphere partition functions of pure 3d Chern-Simons gauge theories.}

\begin{document}
\maketitle
\flushbottom
\allowdisplaybreaks

%%%%%%%%%%%
%%%%%%%%%%%
%%%%%%%%%%%
\section{Introduction}
%%%%%%%%%%%
%%%%%%%%%%%
%%%%%%%%%%%
The holographic interpretation of the entropy of 5d BPS Kerr-Newman black holes \cite{Gutowski:2004yv,Gutowski:2004ez} from the dual field theory point of view
has been an active field of research in the recent past, thanks to the extremization principle of  \cite{Hosseini:2017mds}.  It has been indeed possible to find a microscopic way to count the 
microstates \cite{Choi:2018hmj,Benini:2018ywd}. by extracting them from the superconformal index (SCI) \cite{Romelsberger:2005eg,Kinney:2005ej}.
Further generalizations of these results  have then been obtained \cite{
Honda:2019cio,ArabiArdehali:2019tdm,Kim:2019yrz,Cabo-Bizet:2019osg,Amariti:2019mgp,GonzalezLezcano:2019nca,Lanir:2019abx,Cabo-Bizet:2019eaf,ArabiArdehali:2019orz,Murthy:2020rbd,Cabo-Bizet:2020nkr,Agarwal:2020zwm,Benini:2020gjh,Cabo-Bizet:2020ewf,GonzalezLezcano:2020yeb,Amariti:2020jyx,Amariti:2021ubd,Aharony:2021zkr,Colombo:2021kbb,Choi:2021rxi,Cabo-Bizet:2021plf,Murthy:2022ien,Boruch:2022tno,Mamroud:2022msu}.
It was then realized that it is possible to furnish a field theoretical interpretation of the result in terms of an effective field theory
analysis that follows from the compactification of 4d $SU(N_c)$  SYM on $S^1$ \cite{Cassani:2021fyv,ArabiArdehali:2021nsx}. The analysis is performed by considering  the most general supersymmetric action in 3d and by fixing the coefficients   by a one-loop calculation of the Kaluza-Klein modes on the circle. 
The analysis generalizes the one done in \cite{DiPietro:2014bca} for the ordinary Cardy-limit of the SCI and for the case of  4d $SU(N_c)$ SYM it reproduces the results expected from the matrix model \cite{GonzalezLezcano:2020yeb}.
The EFT  corresponds to the $SU(N_c)_{\pm N_c}$ CS action of an $\mathcal{N}=2$  vector multiplet with further contributions of global CS that can be associated 
to the 4d global anomalies.
On the other hand   the EFT interpretation is less clear when the analysis is performed in the regime of charges that dominates the behavior of the SCI for rational values of the fugacity associated to the rotation parameter \cite{ArabiArdehali:2021nsx}. 
Anyway, from the matrix model calculation, also in this case  a 3d CS theory is expected. Indeed the 3d matrix model corresponds to the one obtained from an  $SU(N_c/C) \times U(1)^{C-1}$ gauge group, with (mixed) CS levels and further contributions that resemble the ones of the global CS discussed in the EFT 
interpretation of the SCI in the BH regime.

It is natural to wonder  how the EFT interpretation generalizes beyond the case of  $su(N_c)$  SYM. A first attempt consists of considering the case of $usp(2n)$ and $so(m)$ gauge algebra, where some results from the matrix model perspective have been obtained in \cite{Amariti:2020jyx}.
In this case for $n=2N_c$ and $m=2N_c+1$ a further question consists of understanding the fate of the S-duality under the Cardy-like limit.
The role of the size of the circle (i.e. the fact that one sums over the whole KK tower) suggests that S-duality should be preserved in the 3d EFT.
This is indeed very similar to the idea pursued in \cite{Aharony:2013dha} for the reduction of 4d dualities to 3d. The finite size effects in the circle reduction there (on the first sheet in the language of \cite{Cassani:2021fyv}) encrypted in the constraints imposed 
by the KK monopole, became crucial in order to construct the 3d EFT preserving the 4d dualities.
This expectation was confirmed from the matrix model calculation  in \cite{Amariti:2020jyx}, restricting to the saddles at vanishing holonomies, the ones that dominate the index in the BH regime.

Physically the matching of the index evaluated on the saddles at vanishing holonomies can be  understood from  the EFT interpretation. First of all in this case the result can be expressed in terms of the 4d trace anomalies, that naturally match across S-dual phases. Second, the less trivial aspect of this matching consists of comparing the contributions from the CS sectors.
The agreement in this case can be reformulated as the fact that S-duality is preserved because the topological sectors, identified by the saddle point holonomies, are equivalent.

However, a full understanding of S-duality in the Cardy-like limit requires to go beyond the case at vanishing holonomies.

In \cite{Amariti:2020jyx} indeed further saddles of the SCI have been studied for the $usp(2N_c)$ case. The behavior of the SCI evaluated on these saddles is generically subleading in the region of charges that reproduces the BH entropy \footnote{With the exception of a saddle  that contributes identically to the one obtained at vanishing holonomies. Such a saddle accounts for the role of the one-form symmetry in the EFT interpretation \cite{Amariti:2021ubd,Cassani:2021fyv}.}. 
On the other hand for the orthogonal case the index has been evaluated so far only for the saddle at vanishing holonomies.
The questions is then if the Cardy-like limit of the SCI of $so(2N_c+1)$ SYM  on these other saddles 
matches the results obtained in \cite{Amariti:2020jyx} for the $usp(2N_c)$ case. Indeed, despite the fact that such saddles are expected to be subleading in the BH regime, they dominate the index in other regions of charges.

In this paper we provide an answer to this question, showing that S-duality relating $so(2N_c+1)$ and $usp(2N_c)$  is fully preserved in the Cardy-like limit of the SCI for small collinear angular momenta. 
In order to provide the complete answer we first study the saddle point equations for $so(2N_c+1)$ SYM, expanding the index at finite $N_c$ in terms of the small angular momenta. Then, we study 
 the behaviour of the index focusing only on the leading terms in the Cardy-like expansion, showing that in various ``physical" regions of charges 
the leading contributions to the index match across the S-dual phases. 
 However, a large $N_c$ limit is required if we stick  to a leading order Cardy-like expansion, to achieve a matching. For this reason, we proceed then to go beyond the leading order and we observe that only after including subleading terms in the expansions S-duality is properly recovered at finite $N_c$.
These last expansions provide also  3d CS partition functions for  topological gauge theories and their evaluation is crucial for our scopes. Indeed by direct evaluation we show that  such CS partition functions vanish on the  saddles that are   subleading at large $N_c$ for any choice of charges. We refer to saddles of this type as perturbatively unstable, because even if they are apparently giving a contribution to the index at leading order in the angular momenta, they vanish once higher order terms in the
expansion are considered.

Summarizing: we find that the SCI of $so(2N_c+1)$ and $usp(2N_c)$ $\mathcal{N}=4$ SYM in the Cardy like limit receives non-vanishing contributions only from a subset of solutions of the saddle point equations. Furthermore, the index expanded in terms of the small collinear angular momenta around such solutions matches among the S-dual theories.

%%%%%%%%%%%
%%%%%%%%%%%
%%%%%%%%%%%
\section{The Cardy-like limit of the SCI of $\mathcal{N}=4$ $usp(2N_c)$ SYM }

\label{secsp}
In this section we overview the results of \cite{Amariti:2020jyx} for the evaluation of the Cardy-like limit of the superconformal index 
for $\mathcal{N}=4$ SYM with gauge algebra
\footnote{Observe that in the following we will always refer to the gauge algebra instead of the gauge group because the superconformal index does not distinguish the global properties of the gauge group.}  $\mathfrak{g} = usp(2N_c)$.
The index corresponds to a matrix integral over the holonomies $u_i$ $i=1,\dots,N_c$ and it can be written in terms of the elliptic Gamma functions as
\begin{equation}
  \begin{split}
    \mathcal{I}^{usp(2N_c)}=&\frac{(p;p)_\infty^{N_c}(q;q)_\infty^{N_c}}{2^{N_c}N_c!}\prod_{a=1}^3\Tilde{\Gamma}(\Delta_a)^{N_c}\int\prod_{i=1}^{N_c}\mathrm{d}u_{i}\frac{\prod_{a=1}^3\prod_{i<j}\Tilde{\Gamma}(\pm  u_{ij}^{(\pm)}+\Delta_a)}{\prod_{i<j}\Tilde{\Gamma}(\pm  u_{ij}^{(\pm)})}\cdot\\
    &\cdot\frac{\prod_{a=1}^3\prod_{i=1}^{N_c}\Tilde{\Gamma}(\pm 2 u_{i}+\Delta_a)}{\prod_{i=1}^{N_c}\Tilde{\Gamma}(\pm 2 u_{i})}.
  \end{split}
  \label{eq:USp_index}
\end{equation}
where $\Delta_{1,2,3}$ are the R-charges of the three adjoints. We refer the reader  to appendix \ref{appA} for the  definition of the  superconformal index 
and to appendix \ref{appB} for the  definitions of the
 elliptic functions and  their asymptotic behavior. 
The index can be also written as an integral of an effective action $S_{\mathrm{eff}}^{usp(2N_c)}$, that in this case is written as
\begin{equation}
    \begin{split}
        S_{\mathrm{eff}}^{usp(2N_c)} &= \sum_{a=1}^3\left(\sum_{i<j}\log\Tilde{\Gamma}\left(\pm u_{ij}^{(\pm)}+\Delta_a\right)+\sum_{i=1}^{N_c}\log\Tilde{\Gamma}\left(\pm 2 u_i+\Delta_a\right)+N_c\log\Tilde{\Gamma}\left(\Delta_a \right)\right)\\
        &+\sum_{i<j}\log\theta_0\left(\pm u_{ij}^{(\pm)}\right)+\sum_{i=1}^{N_c}\log\theta_0\left(\pm 2 u_i\right)+2N_c\log(p;p)_\infty,
    \end{split}
    \label{eq:action_USP}
\end{equation}
such that the matrix integral (\ref{eq:USp_index}) becomes
\begin{equation}
    \mathcal{I}^{usp(2N_c)} = \frac{1}{{2^{N_c}N_c!}} \int\prod_{i=1}^{N_c}\mathrm{d}u_{i}\, e^{ -S_{\mathrm{eff}}^{usp(2N_c)}}
\end{equation}
The next step consists of evaluating the index in the limit $|\tau| \rightarrow 0$ (at fixed $\arg \tau \in (0,\pi)$) restricting to the case 
$\tau =\sigma$ (see \cite{Ardehali:2021irq} for the generalization to $\sigma \neq \tau$).
The evaluation of  the index in this limit corresponds  to a series expansion in $\tau$; such expansion is obtained 
 by  perturbations around  the holonomies that solve the saddle point equations obtained from (\ref{eq:action_USP}).
 As $|\tau| \rightarrow 0$ the saddles will converge to the leading ones, capturing the full behaviour of the index up to exponentially suppressed terms in $|\tau|$.
The saddle point equations are
\begin{equation}
    \begin{split}
        \sum_{a=1}^3\Bigg[\sum_{\substack{j=1\\j\neq k}}^{N_c}\bigg( \!\! B_2\{u_{ij}^{(\pm)} + \Delta_a\} \!-\! B_2\{-u_{ij}^{(\pm)} + \Delta_a\} \!\! \bigg) + B_2\{2u_i +\Delta_a\} \!-\! B_2\{- 2u_i +\Delta_a\} \Bigg] \! \!=\! 0.
    \end{split}
    \label{eq:SP_saddle_eq}
\end{equation}
The analysis of the solutions of these equations and the expansion of the index has been performed in \cite{Amariti:2020jyx}.
In the following we review the results. It has been observed that 
the index receives contributions from two families of saddle points
and that the final sum over such saddles can be written as
\begin{equation}
\label{eq:usp_index_cardy}
    \mathcal{I}^{usp(2N_c)} =\!\! \sum_{L=0}^{\lfloor \frac{N_c - 1}{2} \rfloor} \!\!\textcolor{red}{2\mathcal{I}_{L=0,N_c - L = 1/2}^{usp(2N_c)} } + \textcolor{blue}{\mathcal{I}_{L=0,L = 1/2, N_c - 2L = 1/4}^{usp(2N_c)} } + \left( {\color{ufogreen}\mathcal{I}_{N_c/2=0,N_c/2 = 1/2}^{usp(2N_c)} } \;\; \mathrm{if\; N_c  \; even} \right)\!.
\end{equation}
Each of the families has a distinct leading saddle point which dominates in a specific region of charges.
\begin{itemize}
    \item  The \textcolor{red}{first} family is constituted of saddles with $L$ holonomies at $u_i = 0$ and $K \equiv N_c - L$ holonomies at $u_i = 1/2$. Such saddles are paired by  the relation $\mathcal{I}_{L,N_c - L} = \mathcal{I}_{N_c - L, L} $. For this reason it is convenient to count them starting from $L = 0$ up to $\lfloor \frac{N_c - 1}{2} \rfloor$ with a degeneracy factor 2.
 Their contribution to the index has been studied in \cite{Amariti:2020jyx}, here we only report the result.

The saddle point and the effective action emerging near the saddle as $|\tau|\rightarrow0$ are
\begin{equation}
    \hat{\mathbf{u}}=
    \begin{cases}
    \bar{u}_j = v_j\tau \hspace{4.8cm}  j = 1,...,L\\
    \frac{1}{2} + \bar{u}_{L+r} \equiv \frac{1}{2} + \bar{w}_{r} = \frac{1}{2} + w_{r}\tau \hspace{1cm}  r = 1,...,N_c - L.
    \end{cases}
    \label{eq:USp0half_ansatz}
\end{equation}

\begin{equation}
  \begin{split}
    &S_{L,K}^{usp(2N_c)} = \\
    &=- \frac{2\pi i}{\tau^2} \left(\eta_1(L+1-K) + K\eta_2\right)\sum_{i=1}^L \bar{u}_i^2 -\frac{2\pi i}{\tau^2} \left(\eta_1(K + 1 - L) + L\eta_2\right)\sum_{r=1}^K \bar{w}_r^2 \\
    &+\sum_{i < j}^L \log \left[ 2\sin \left({\pm \frac{\pi \bar{u}_{ij}^{(\pm)}}{\tau}} \right) \right] + \sum_{i=1}^{L}  \log \left[2\sin \left({\pm\frac{2\pi \bar{u}_i}{\tau}}\right)\right]  \\
    &+ \sum_{r < s}^K  \log \left[ 2\sin\left({\pm\frac{\pi \bar{w}_{rs}^{(\pm)}}{\tau}}\right)\right] +\sum_{r=1}^{K}  \log\left[2 \sin\left({\pm \frac{2\pi \bar{w}_r}{\tau}} \right) \right]\\
    & - \frac{i\pi}{\tau^2}\left( 2(L\! -\! K)^2 + N_c \right)\prod_{a=1}^3 \left(\{\Delta_a\}_\tau\! -\!\frac{1+\eta_1}{2}\right) \! -\! \frac{i\pi}{\tau^2}LK  \prod_{a=1}^3\left(\!\{2\Delta_a\}_\tau \!-\! \frac{1+\eta_2}{2}\right)\\
    &+ i \pi\left( \frac{(6-5\eta_1)\left(2(L-K)^2+N_c\right)}{12} \! +\! \frac{(12 - 5\eta_2)LK}{3} -N_c^2\!\right) \!-\!N_c\log(\tau) +\mathcal{O}(\tau),
  \end{split}
    \label{eq:USp0half_eff_S}
\end{equation}
where $\eta_1 = \pm 1$ and $\eta_2 = \pm 1$ define different chambers for the chemical potentials, satisfying
\begin{equation}
    \sum_{a = 1} ^ 3 \{ I \Delta_a \}_\tau = 2\tau + \frac{3 + \eta_I}{2}
\end{equation}
These constraints arise as a consequence of the constraint
\begin{equation}
    \prod_{a = 1} ^ 3 y_a = p q \implies \Delta_1 + \Delta_2 + \Delta_3 - 2\tau \in \mathbb{Z},
\end{equation}
together with the requirement that $\Delta_a\not\rightarrow 0,2$.
The reduction over the thermal $S^1$ with length $\beta$ in the Cardy-like limit $\tau \sim \beta$ produces 3d pure CS partition functions on $S^3$ after the integration of the massive KK modes on $S^1$. The original $usp(2N_c)$ gauge algebra is broken down to $usp(2L)_{k_1}\times usp(2K)_{k_2}$. This can be read off directly from the effective action, as it is reflected in the logarithmic terms in (\ref{eq:USp0half_eff_S}), defining the measure of the CS partition function, upon exploiting property (\ref{hyp_sin}) (with $\omega_1 = \omega_2 = i$) for the hyperbolic gamma functions. The CS levels can be identified by recalling the expression for a pure 3d CS partition function on $S^3$ with $usp(2m)_k$ gauge algebra
\begin{equation}
  Z_{S^3}^{usp(2m)} = \frac{e ^ { i\pi m^2 }}{|2^{m} m!| } \int \prod_{i=1}^{m}
d \sigma_i
e^{- i \pi 2 k \sigma_i^2 }
\prod_{\alpha \in \Delta_{+}} 4 \sinh( \pm \pi \alpha(\sigma))).
\end{equation}
Upon making the CS effective action apparent, through the change of variables $\bar{u}_j = - i \sigma_j \tau$, the CS levels are $k_{1,2} = - C_{1,2}$, with $-\frac{2\pi i}{\tau^2}C_{1,2}$ being the coefficients of the quadratic terms in (\ref{eq:USp0half_eff_S}).

All in all, the contribution to the SCI coming from the $(L,K)$ saddle point of this family is, up to exponentially suppressed corrections $\sim \mathcal{O}(e^{ -\frac{1}{|\tau|}})$ in the Cardy-like limit,

\begin{equation}
    \mathcal{I}_{L,K}^{usp(2N_c)} = \tau^{N_c}e^{-\frac{i \pi \left(2(L-K)^2 + N_c\right)}{2}}e^{-2i\pi LK}\mathcal{I}_0 Z_{S^3}^{usp(2L)_{k_1}}Z_{S^3}^{usp(2K)_{k_2}},
\end{equation}

where
\begin{equation}
\begin{split}
\log\mathcal{I}_0 \equiv & - \frac{i\pi}{\tau^2}\left( 2(L - K)^2 + N_c \right)\prod_{a=1}^3\left(\{\Delta_a\}_\tau-\frac{1+\eta_1}{2}\right)  +\\
 & - \frac{i\pi}{\tau^2}LK   \prod_{a=1}^3\left(\{2\Delta_a\}_\tau - \frac{1+\eta_2}{2}\right) -N_c\log(\tau) +\\
    &+i \pi\left( \frac{(6-5\eta_1)\left(2(L-K)^2+N_c\right)}{12} + \frac{(6 - 5\eta_2)LK}{3} -(L^2 + K^2)\right).
\end{split}
\end{equation}
and the CS levels for the 3d pure CS theories partition functions on $S^3$ are
\begin{equation}
\label{eq:levels_usp0half}
    \begin{cases}
    k_1 = - ((L + 1 - K)\eta_1+K\eta_2)\\
    k_2 = - ((K + 1 - L)\eta_1 + L\eta_2).
    \end{cases}
\end{equation}

\item The \textcolor{blue}{second} family is described by saddles with $L$ holonomies at $u_i = 0$, $L$ at $u_i = 1/2$ and $K = N_c - 2L$ at $u_i = 1/4$ and $L$ ranging between zero and $\lfloor \frac{N_c - 1}{2} \rfloor$.
The original gauge algebra is broken by these vacua, with breaking pattern $usp(2N_c)\rightarrow usp(2L)_{k_1}\times usp(2L)_{k_2}\times su(K)_{k_3}\times u(1)_{k_4}$ and a pure 3d CS partition function emerges. The contribution to the SCI coming from these saddles is 
\begin{equation}
    \mathcal{I}_{L,L,K}^{usp(2N_c)} = \tau^N e^{-\frac{i \pi (N_c + 4L^2 + K(K-1)) }{2}} \mathcal{I}_0  Z_{S^3}^{usp(2L)_{k_1}}Z_{S^3}^{usp(2L)_{k_2}}Z_{S^3}^{su(K)_{k_3}}Z_{S^3}^{u(1)_{k_4}}
    \label{eq:index_final_0_1_2_1_4}
\end{equation}
with
\begin{equation}
\begin{split}
\log\mathcal{I}_0 =& \frac{i\pi(2L-K)}{\tau^2}\prod_{a=1}^3\left(\{\Delta_a\}_\tau - \frac{1+\eta_1}{2}\right) - \frac{i\pi LK}{4\tau^2}\prod_{a=1}^3\left(\{4\Delta_a\}_\tau - \frac{1+\eta_4}{2}\right) \\
-&\frac{i\pi((2L - K)^2 + K)}{4\tau^2}\prod_{a=1}^3\left(\{2\Delta_a\}_\tau - \frac{1+\eta_2}{2}\right) - i \pi (4L^2 +K^2) \\
    +&  \frac{i\pi(6-5\eta_1)(2L \!-\! K)}{12} \!+\! \frac{i\pi(12\!-\!5\eta_2)(2L \!-\! K)^2 + K)}{12} \!+\! \frac{i\pi(12 \!-\! 5\eta_4)LK}{3},
\end{split}
\end{equation}
the $\eta_i$ defined similarly as before and 
\begin{equation}
    \begin{cases}
    k_1 = -\frac{1}{2}\big(2\eta_1 + (2L - K)\eta_2 + K\eta_4\big)\\
    k_2 = -\frac{1}{2}\big(2\eta_1 + (2L - K)\eta_2 + K\eta_4\big)\\
    k_3 = -\big(-2\eta_1 + (K - 2L + 2)\eta_2 + 2L\eta_4 \big)\\
    k_4 = -2\big(-(K + 1)\eta_1 + (K + 1 - L)\eta_2 + L\eta_4 \big).
    \end{cases}
\end{equation}
\item When $N_c$ is even there is also a self-paired \textcolor{ufogreen}{saddle} with $N_c/2$ holonomies at $0$ and $1/2$; such saddle represents a limiting case of the other two families discussed above. 
\end{itemize}

Summarising, the index of $\mathcal{N} = 4$ $usp(2N_c)$ SYM receives contributions from  $N_c + 1$ distinct saddle points, divided in two families. Employing the pairing degeneracy discussed above, the saddles of the two families can be combined naturally into one, parameterised by $\mathcal{I}_j$, with $j=0,...,N_c - 1$ and defined as follows:

\begin{equation}
\begin{cases}
    \mathcal{I}_j \equiv \textcolor{blue}{\mathcal{I}_{j,j,N_c - 2j}^{usp(2N_c)} } & \quad 0 \leq j \leq \lfloor \frac{N_c}{2} \rfloor \\
    \mathcal{I}_j \equiv \textcolor{red}{2\mathcal{I}_{j,N_c - j}^{usp(2N_c)} } & \quad \lceil \frac{N_c}{2} \rceil \leq j \leq N_c.
\end{cases}
\end{equation}

The limiting case $j = \frac{N_c}{2}$ is common to both families and connects them, resulting in a well ordered distribution of saddles shown in Figure \ref{fig:usp saddles}.

\subsection{Explicit evaluation}
In this section, we perform a complete analysis of the contributions to the SCI from each saddle. At first, we will focus on the leading order Cardy-like limit, such to identify the dominant saddle points in the regions of charges denoted as 
\emph{physical}  in \cite{ArabiArdehali:2021nsx}.
In our language these correspond to the choices $\eta_1 = -\eta_2 = \pm 1$ which reduce to  
the cases discussed in \cite{ArabiArdehali:2021nsx}
when $\{\Delta_a \} = 1/3, 2/3$.

We show that a leading order analysis in $1/|\tau|^2$, while being enough to determine the dominant saddle points in each region of charges, can miss physical properties of these vacua such as S-duality and possible perturbartive instabilities that emerge in the calculation at  subleading orders in $|\tau|$  and that are encoded in the three-sphere CS partition functions. For this reason, we claim that a complete expansion beyond the leading order in the Cardy-like limit is necessary to achieve a physically reliable result. 

The leading order $ 1/|\tau|^2$ competition between the saddles in each family is determined by a parabola.
For the \textcolor{red}{first} family we have
\begin{equation}
\label{eq:parabola_USp0half}
-\frac{\tau^2}{i \pi} S_{L,N_c - L}^{usp(2N_c)} =  \left(2(N_c - 2L)^2 + N_c \right)\alpha_{1} - L(N_c - L)\alpha_{2},
\end{equation}
where we defined
\begin{equation}
\label{alphabeta}
    \alpha_{I} \equiv \prod_{a = 1} ^ {3}\left(\{I \Delta_a\}_\tau - \frac{1 + \eta_I}{2} \right) \quad.
\end{equation}
We can  determine the dominant saddle point in both chambers  $\eta_1 = -\eta_2 = 
\eta_4 = \mp 1$. The net effect of switching from the first to the second region is to change the concavity of the parabola, switching from a M-shaped effective potential to a W-shaped one in the language of \cite{ArabiArdehali:2019tdm}.

The vertex of (\ref{eq:parabola_USp0half}) sits at 
\begin{equation}
    L = \frac{N_c}{2} 
\end{equation}
as expected due to the pairing between the $L$ and the $N_c - L$ saddles. Thus, the leading saddle is either the one closer to $N_c/2$ or the saddle with $N_c$ holonomies at zero, depending on the chamber of the chemical potentials we are in.

Analogously, for the \textcolor{blue}{second} family we find that the vertex of the corresponding parabola sits at
\begin{equation}
\label{eq:vertex_USpMix}
    L = \frac{N_c}{4} -  \frac {8 \alpha_1 - \alpha_2 }{2 (8\alpha_2 - \alpha_4)}
\end{equation}

\begin{figure}[h]
\centering
\begin{subfigure}{.5\textwidth}
  \centering
  \includegraphics[width=\linewidth]{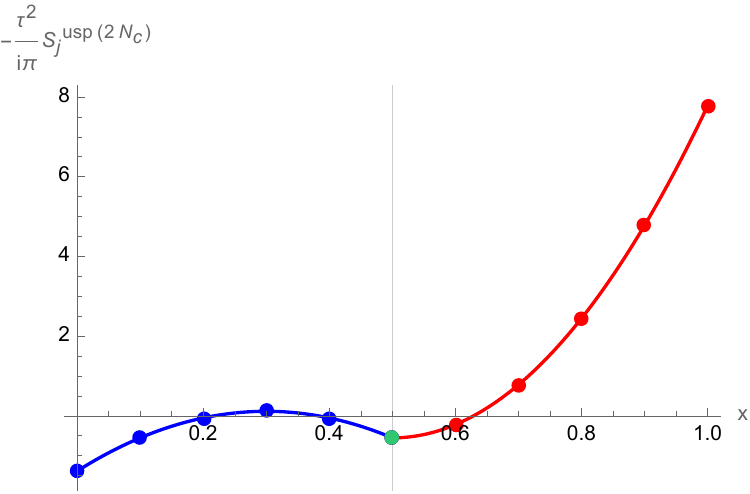}
\end{subfigure}
\begin{subfigure}{.49\textwidth}
  \centering
  \includegraphics[width=\linewidth]{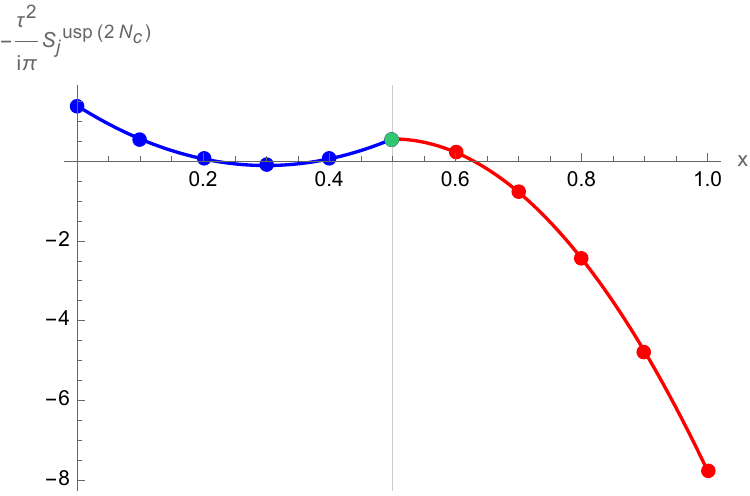}
\end{subfigure}
\caption{\footnotesize{Behaviour of the saddle points for $usp(2N_c)$ as $x \coloneqq j/N_c$ ranges from 0 to 1 and $N_c = 10$. On the left: $\{\Delta_a\} = 1/3$; on the right: $\{\Delta_a\} = 2/3$.}}
\label{fig:usp saddles}

\end{figure}

In the ``physical" regions the relation $ \frac {8 \alpha_1 - \alpha_2}{2 (8 \alpha_2- \alpha_4)} < 0 $ always holds and it allows us to conclude that the leading saddle for this family is either the one with $N_c$ holonomies at $u_i = 1/4$ or the saddle closer to the vertex (\ref{eq:vertex_USpMix}) defined by some $L$ (say $L^*$) by symmetry reasons. However, we notice that the two parabolas describing the two families of saddles have opposite concavities, thus depending on the region of the chemical potentials the leading saddle is either the one where the $N_c$ holonomies sitting at zero  dominate on the $L^*$ saddle of the second family, or the one with $N_c$ holonomies at $1/4$ (in the W winged shaped potential) as shown in (Fig. \ref{fig:usp saddles}).
 Borrowing again the terminology of \cite{ArabiArdehali:2019tdm}, we refer to the choice where the vanishing holonomies dominate as the  M-wing, while the region where the non-vanishing holonomies dominate is referred to the W-wing.
The first case corresponds to the choice $\eta_1 = -\eta_2 = - 1$, while the second case corresponds to  $\eta_1 = -\eta_2 =  1$.

%%%%%%%%%%%
%%%%%%%%%%%
%%%%%%%%%%%

%%%%%%%%%%%
%%%%%%%%%%%
%%%%%%%%%%%
\section{The Cardy-like limit of the SCI of $\mathcal{N} \!=\! 4$ of $so(2N_c + 1)$ SYM}
\label{sec:SO_saddles}
%%%%%%%%%%%
%%%%%%%%%%%
%%%%%%%%%%%

In this section we  focus on the Cardy-like limit evaluation of the SCI for $4d$ $\mathcal{N} = 4$ SYM with $so(2N_c + 1)$ gauge algebra, determining the general structure of the saddle points. The index is given by
\begin{equation}
  \begin{split}
    \mathcal{I}^{so(2N_c+1)}=&\frac{(p;p)_\infty^{N_c}(q;q)_\infty^{N_c}}{2^{N_c}N_c!}\prod_{a=1}^3\Tilde{\Gamma}(\Delta_a)^{N_c}\int\prod_{i=1}^{N_c}\mathrm{d}u_{i}\frac{\prod_{a=1}^3\prod_{i<j}\Tilde{\Gamma}(\pm u_{ij}^{(\pm)}+\Delta_a)}{\prod_{i<j}\Tilde{\Gamma}(\pm u_{ij}^{(\pm)})}\cdot\\
    &\cdot\frac{\prod_{a=1}^3\prod_{i=1}^{N_c}\Tilde{\Gamma}(\pm u_{i}+\Delta_a)}{\prod_{i=1}^{N_c}\Tilde{\Gamma}(\pm u_{i})}.
  \end{split}
  \label{eq:SO_index}
\end{equation}
We define the effective action
\begin{equation}
    \begin{split}
        S_{\mathrm{eff}}^{so(2N_c+1)} &= \sum_{a=1}^3\left(\sum_{i<j}\log\Tilde{\Gamma}\left(\pm u_{ij}^{(\pm)}+\Delta_a\right)+\sum_{i=1}^{N_c}\log\Tilde{\Gamma}\left(\pm u_i+\Delta_a\right)+N_c\log\Tilde{\Gamma}\left(\Delta_a \right)\right)\\
        &+\sum_{i<j}\log\theta_0\left(\pm u_{ij}^{(\pm)}\right)+\sum_{i=1}^{N_c}\log\theta_0\left(\pm u_i\right)+2N_c\log(p;p)_\infty,
    \end{split}
    \label{eq:action_SO}
\end{equation}
such that the index is
\begin{equation}
     \mathcal{I}^{so(2N_c+1)} = \frac{1}{{2^{N_c}N_c!}} \int\prod_{i=1}^{N_c}\mathrm{d}u_{i}\, e^{ -S_{\mathrm{eff}}^{so(2N_c+1)}}
\end{equation}

General solutions to the saddle point equations beyond the leading order Cardy-like limit can be found by first focusing on the leading term in the Cardy-like limit and then by expanding around those solutions accordingly, following the strategy of \cite{GonzalezLezcano:2020yeb}. As $|\tau| \rightarrow 0$ the saddles will converge to the leading ones, capturing the full behaviour of the index up to exponentially suppressed terms in $|\tau|$.\\
The saddle point equations are
\begin{equation}
    \begin{split}
    \sum_{a=1}^3\Bigg[\sum_{\substack{j=1\\j\neq k}}^{N_c}\bigg(\!B_2\{u_{ij}^{(\pm)} \!+\! \Delta_a\} \!-\!\! B_2\{-u_{ij}^{(\pm)} + \Delta_a\}\!\bigg) + B_2\{u_i +\Delta_a\} \!-\! B_2\{- u_i +\Delta_a\} \Bigg] \!=\! 0.
    \end{split}
    \label{eq:SO_saddle_eq}
\end{equation}

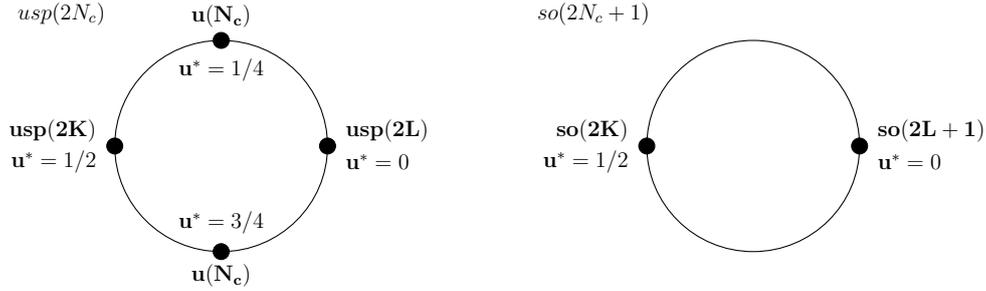
\begin{figure}[h]
\centering
    \begin{tikzpicture}[scale=.7, transform shape]
    \draw node at (-3,2.5) {$usp(2N_c)$};
      \coordinate (A) at (0,0);
      \draw (A) circle(2cm);
      \filldraw[fill=black] (2,0) circle (0.15);
      \draw node[right] at (2.2,0.3) {$\mathbf{usp(2L)}$};
      \draw node[right] at (2.2,-0.3) {$\mathbf{u}^* = 0$};
      \filldraw[fill=black] (0,2) circle (0.15);
      \draw node[above] at (0,2.1) {$\mathbf{u(N_c)}$};
      \draw node[below] at (0,1.8) {$\mathbf{u}^* = 1/4$};
      \filldraw[fill=black] (-2,0) circle (0.15);
      \draw node[left] at (-2.2,0.3) {$\mathbf{usp(2K)}$};
      \draw node[left] at (-2.2,-0.3) {$\mathbf{u}^* = 1/2$};
      \filldraw[fill=black] (0,-2) circle (0.15);
      \draw node[below] at (0,-2.1) {$\mathbf{u(N_c)}$};
      \draw node[above] at (0,-1.8) {$\mathbf{u}^* = 3/4$};
      \draw node at (7,2.5) {$so(2N_c + 1)$};
      \coordinate (A) at (10,0);
      \draw (A) circle(2cm);
      \filldraw[fill=black] (12,0) circle (0.15);
      \draw node[right] at (12.2,0.3) {$\mathbf{so(2L+1)}$};
      \draw node[right] at (12.2,-0.3) {$\mathbf{u}^* = 0$};
      \filldraw[fill=black] (8,0) circle (0.15);
      \draw node[left] at (7.8,0.3) {$\mathbf{so(2K)}$};
      \draw node[left] at (7.8,-0.3) {$\mathbf{u}^* = 1/2$};
    \end{tikzpicture}
        \caption{\footnotesize{Vacua distribution together with the corresponding group of the effective CS theory.}
        }
        \label{fig:1}
\end{figure}

The absence of a factor 2 in the $\pm u_i$ roots of $so(2N_c + 1)$ with respect to the ones of $usp(2N_c)$ plays a crucial role in the behaviour of the structure of the saddles, leading to a rather different behaviour than the ones of the symplectic case, as showed in  Figure \ref{fig:1}
\footnote{Observe that the gauge symmetry breaking pattern is reminiscent of the one dictated by the split of an orientifold $O4$ plane under T-duality along a compact direction. It would be interesting to investigate further on this relation.}.
We found that the solution with $N_c$ holonomies at zero, already studied in \cite{Amariti:2020jyx}, lies inside a
more general family of saddles parameterised by $L = 0, N_c$, which counts the number of holonomies set to zero. The general saddle point is of the form $(L,N_c-L)$, with $L$ holonomies at zero and $N_c - L$ at $1/2$.
As opposed to the symplectic case, there is no pairing between the $L$ and $L' = N_c - L$ saddles.

 The saddle point beyond the leading order in the Cardy-like limit and the corresponding subleading contributions to the index are then obtained by expanding around the leading saddles.
 
\subsection{$L$ holonomies at $u_i = 0$, $K = N_c - L$ holonomies at $u_i = 1/2$}
We make the following ansatz for the general saddle point:

\begin{equation}
    \hat{\mathbf{u}}=
    \begin{cases}
    \bar{u}_j = v_j\tau \hspace{4.8cm}  j = 1,...,L\\
    \frac{1}{2} + \bar{u}_{L+r} \equiv \frac{1}{2} + \bar{w}_{r} = \frac{1}{2} + w_{r}\tau \hspace{1cm}  r = 1,...,K. 
    \end{cases}
    \label{eq:SO_ansatz_0_1_2}
\end{equation}
Then, expanding the effective action near $\hat{\mathbf{u}}$ for $|\tau| \rightarrow 0$ we obtain

    \begin{eqnarray}
        \label{eq:SO_action_0_1_2}
       && S_{L,K}^{so(2N_c + 1)} = - \frac{i\pi}{\tau^2} (2(L - K) - 1)\eta_1 +  2K\eta_2) \sum_{i=1}^L\bar{u}_i^2 + \sum_{i<j}^L \log \left( 2\sin \left( \frac{ \pm \pi \bar{u}_{ij}^{(\pm)}}{\tau}\right )\right) + \nonumber \\   
       && + \sum_{i=1}^L \log \left(2\sin \left( \frac{ \pm \pi \bar{u}_i}{\tau} \right) \right) - N_c\log(\tau) + \nonumber\\
       && - \frac{i\pi}{\tau^2} ((2(K - L) - 3)\eta_1 + (2L + 1)\eta_2) \sum_{r=1}^K \bar{w}_r^2 + 
       \sum_{r<s}^K \log \left(2\sin \left(\frac{ \pm \pi \bar{w}_{rs}^{(\pm)}}{\tau} \right) \right) + \nonumber \\
       && - \frac{i\pi}{\tau^2} \left(2(L - K)^2 + L - 3K\right) \prod_{a=1}^3 \left(\{\Delta_a\}_\tau - \frac{1 + \eta_1}{2}\right) \nonumber\\
       && - \frac{i\pi}{2\tau^2} \left(K (1 + 2L) \right) \prod_{a=1}^3 \left(\{2\Delta_a\}_\tau - \frac{1 + \eta_2}{2} \right) +  \nonumber \\ 
       && + i \pi \left(\frac{(6 - 5\eta_1)\left(2(L - K)^2 + L - 3K \right)}{12} + \frac{(12 - 5\eta_2)K(1 + 2L)}{6} - N_c^2 \right),
    \end{eqnarray}
where  again $\eta_1 = \pm 1$ and $\eta_2 = \pm 1$.
The action (\ref{eq:SO_action_0_1_2}) is manifestly not invariant under $L \leftrightarrow K$, differently to the symplectic case. Upon changing variables $\bar{u}_j = - i \sigma_j \tau$, we can read off  the three sphere partition function a 3d pure CS theory.
Such CS theories arise by expanding the holonomies around the  $ u_i = 0 $ and $ u_i = 1/2 $ vacua, and they give rise to an odd and even rank orthogonal gauge  group respectively.
We obtain the partition function of an $s(o(2L+1)_{k_1} \times o(2K)_{k_2})$ pure CS theory  with CS levels
\begin{equation}
\label{eq:levels_so0half}
\begin{cases}
k_1 = - (2(L - K) - 1)\eta_1 - 2K\eta_2 \\
k_2 = - (2(K - L) - 3)\eta_1 - (2L + 1)\eta_2
\end{cases}
\end{equation}
The index is then
\begin{equation}
\label{eq:index_so_final}
    \mathcal{I} = \sum_{L = 0} ^ {N_c} \mathcal{I}_{L, N_c - L}, \quad
\text{where} \quad
 \mathcal{I}_{L,K} = \tau^{N_c} 
 e^{-\frac{i \pi((2K - 1) K + (2L + 1) L)}{2}} \mathcal{I}_0 Z_{S^3}^{s(o(2L+1)_{k_1} \times o(2K)_{k_2})}
\end{equation}
and 
\begin{equation}
\begin{split}
    \log\mathcal{I}_0 &= -\frac{i\pi}{\tau^2} \left(2(L-K)^2 +L-3K\right)\prod_{a=1}^3\left(\{\Delta_a\}_\tau-\frac{1+\eta_1}{2}\right)\\
       &- \frac{i\pi}{2\tau^2} \left(K(1+2L)\right)\prod_{a=1}^3\left(\{2\Delta_a\}_\tau-\frac{1+\eta_2}{2}\right) -N_c\log(\tau) +  \\ 
       & +i \pi \left(\frac{(6 - 5\eta_1)\left(2(L-K)^2 +L-3K\right)}{12} + \frac{(12 - 5\eta_2)K(1+2L)}{6} - N_c^2 \right).
\end{split}
\end{equation}

\subsection{General behaviour of the saddles}
Again, the dominant saddle point in the Cardy-like limit depends on the region of chemical potentials we are in. To identify the leading saddle it is enough to focus on the leading order term. The behaviour of the saddles is determined by a second degree polynomial in $L \in [0,N_c]$.
\begin{equation}
    \begin{split}
   -\frac{\tau^2}{i \pi} S_{L,N_c - L}^{so(2N_c + 1)} =   \left(2 ( N_c - 2L )^2 + 4L - 3N_c \right) \alpha_1 + \frac{1}{2} \left( (N_c - L) (1 + 2L) \right) \alpha_{2}.
    \end{split}
    \label{eq:SO exponent}
\end{equation}
where $\alpha_{1,2}$ are defined in (\ref{alphabeta})
The parabola has a vertex in 
\begin{equation}
    L = \frac{2N_c - 1}{4}
\end{equation}
independently of the chemical potentials. 

Thus, it follows that the dominant saddle is either the one closer to the vertex with $L = \left \lfloor \frac{N_c}{2} \right \rfloor$, since $L$ must be integer, or the saddle with  $L = N_c$ at the extremum of the domain of the parabola, depending on the region of chemical potentials we are considering. The saddle with $L=0$ is penalised, due to the vertex being closer to zero than to $N_c$; only in the large $N_c$ limit we expect to recover a pairing between the $L$ and the $L'=N_c - L$ saddles as the symmetry axis of the parabola goes to $N_c/2$. In the ``physical" regions we are in the M-wing or in the W-wing. In the first case the dominant saddle point is the one with $N_c$ holonomies at zero, while in the second case the dominant saddle is the one with $L = \left \lfloor \frac{N_c}{2} \right \rfloor$. 

Summarizing, the M-wing is dominated by vanishing holonomies, while the saddle with  $L = \left \lfloor \frac{N_c}{2} \right \rfloor$ holonomies at $u_i = 0$ and the remaining at $u_i = 1/2$ dominates the W-wing.

\begin{figure}[h]
\centering
\begin{subfigure}{.5\textwidth}
  \centering
  \includegraphics[width=\linewidth]{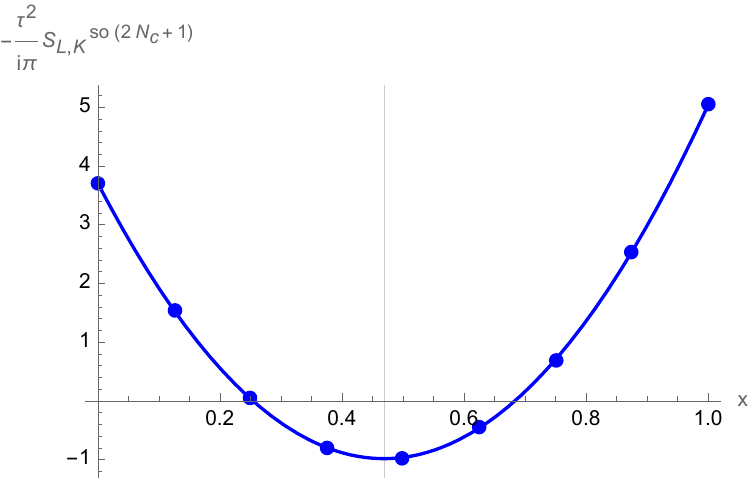}
\end{subfigure}
\begin{subfigure}{.49\textwidth}
  \centering
  \includegraphics[width=\linewidth]{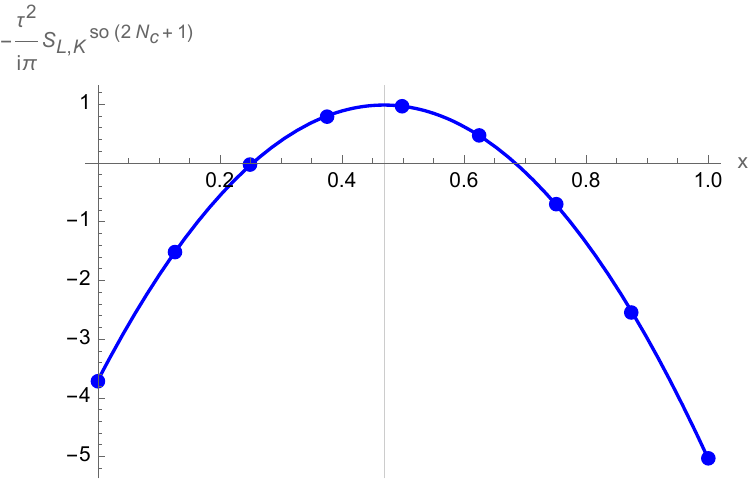}
\end{subfigure}
\caption{\footnotesize{The behaviour of (\ref{eq:SO exponent}) as $x \coloneqq L/N_c$ ranges from 0 to 1. The vertical line passes through the vertex of the parabola. For presentation purposes we plotted the case with $N_c = 8$. On the left: $\{\Delta_a\} = 1/3$; on the right: $\{\Delta_a\} = 2/3$.}}
\label{fig:test}
\end{figure}

%%%%%%%%%%%
%%%%%%%%%%%
%%%%%%%%%%%
\section{S-duality}
%%%%%%%%%%%
%%%%%%%%%%%
%%%%%%%%%%%

In this section we study the fate of 4d S-duality in the Cardy-like limit of the SCI. We start by matching the $1/|\tau|^2$ leading order contribution to the index when all the holonomies are vanishing in both the symplectic and orthogonal case. This saddle dominates the index in the 
M-wing of the potential and it reproduces the  entropy function of the would be holographic dual black hole. Then we match the leading contributions  in the region of parameter where the index is in the W-wing.

Then we consider the fate of S-duality also in presence of subleading contribution in $|\tau|$. As discussed above  only few saddles survive for both $usp(2N_c)$ and $so(2N_c+1)$.
These saddles are exactly  the ones that dominates the index in the M-wing and in the W-wing.
In the case of the M-wing the full matching was discussed in \cite{Amariti:2020jyx}. In the W-wing 
we show here that S-duality is preserved because of a non-trivial identity among the pure CS partition functions.

%%%%%%%%%%%
\subsection{S-duality at the leading order}
%%%%%%%%%%%

We begin our analysis by focusing on the leading $1/|\tau|^2$ order expansion of the index. The dominant contributions to the index in each region $\eta_1 = -\eta_2 = \mp 1 $ have been identified in the previous sections and read
\begin{itemize}
    \item M-wing ($\eta_1 = - 1$): the dominant contribution in the orthogonal case is achieved for vanishing holonomies. The symplectic theory is dominated by the same configuration of holonomies but the contribution is doubled due to the pairing between the saddles at $N_c$ holonomies at $u_i = 0$ and at $N_c$ holonomies at $u_i = \frac{1}{2}$. As discussed in \cite{Amariti:2020jyx},  the factor 2 degeneracy, understood as the presence of a $\mathbb{Z}_2$ global 1-form symmetry, is not apparent in the orthogonal theory at this order, for any finite $N_c$, and only once subleading corrections 
    in $|\tau|$ are included such factor can be recovered. 
    \item W-wing ($\eta_1 = 1$): The symplectic theory is dominated by the saddle point with $N_c$ holonomies at $u_i = \frac{1}{4}$, while for the orthogonal case the dominant contribution arise when $L = \lfloor \frac{N_c}{2}\rfloor$ holonomies sits at $u_i = 0$ and the remaining ones at $u_i = \frac{1}{2}$.
\end{itemize}

At this level the expectation is that S-duality manifests as a matching between the dominant saddles. A natural question regards the role of the regions of chemical potentials. When an EFT interpretation of the underlying 3d pure CS theory is understood, the matching between the saddles is actually constrained by S-duality independently of the specific region of $\eta_I$ we sits in, because the topological sectors identified by the holonomies are  equivalent. This is indeed the case for the saddle at vanishing holonomies for which an EFT interpretation for the CS terms can be recovered.

To be more explicit, one can readily observe that
\begin{equation}
\begin{split}
   \mathcal{I}_{N_c,0}^{usp(2N_c)} = \mathcal{I}_{N_c,0}^{so(2N_c + 1)}  =  \exp\left[-\frac{i\pi N_c(2N_c+1)}{\tau^2} \prod_{a=1}^3\left(\{\Delta_a\} - \frac{1+\eta_1}{2}\right)\right]
\end{split}
\end{equation}
which holds for any value of $\eta_I$ and it thus persists independently of the specific wing we are in.

Notice however that by sticking at order $1/|\tau|^2$ a proper matching can be achieved only considering the large $N_c$ limit, when the reflexive symmetry $L\leftrightarrow K$ between the saddles is recovered also in the orthogonal case. In fact, for $N_c \rightarrow \infty$ we get
\begin{equation}
\begin{split}
    &\mathcal{I}_{N_c,0}^{usp(2N_c)} + \mathcal{I}_{0,N_c}^{usp(2N_c)} =  \mathcal{I}_{N_c,0}^{so(2N_c + 1)} + \mathcal{I}_{0,N_c}^{so(2N_c + 1)} = \\
    =&\, 2\exp \left[-\frac{i\pi N_c(2N_c+1)}{\tau^2} \prod_{a=1}^3\left(\{\Delta_a\} - \frac{1+\eta_1}{2}\right)\right]
\end{split}
\end{equation}

The same argument cannot be employed for the saddles dominating the W-wing as the EFT interpretation is less clear. However, also for these saddles the matching extends to any region of $\{\Delta_a\}$ at least for the leading order in $|\tau|$.

Indeed,

$$
   \mathcal{I}_{0,0,N_c}^{usp(2N_c)} = \mathcal{I}_{\lfloor \frac{N_c}{2} \rfloor, \lceil \frac{N_c}{2} \rceil } ^{so(2N_c + 1)}  =
$$
\begin{equation}
   = \exp \left [\frac{i \pi N_c }{\tau^2}\prod_{a=1}^{3} \left(\{\Delta_a\} -\frac{1 + \eta_1}{2}\right) - \frac{i \pi (N_c^2+N_c) }{4\tau^2}\prod_{a=1}^{3}\left( \{2\Delta_a\} - \frac{1 + \eta_2}{2}\right) \right]
\end{equation}

At this level of the discussion the fate of S-duality on the other saddles is unclear. Indeed we did not find a matching among the indices expanded 
around such saddles at leading order in $1/|\tau|^2$. The situation is clarified by taking into account the complete expansion in $|\tau|$, as we will show in the next sub-section.

\subsection{Beyond the leading order}

The 3d CS partition function on $S^3$ is
\begin{align}
Z_{\mathfrak{g}} = \frac{1}{|W| } \int \prod_{i=1}^{\text{rk}\mathfrak{g}}
d \sigma_i
e^{\frac{i \pi k \sigma_i^2 }{\omega_1 \omega_2} }
\prod_{\alpha(\sigma)} \Gamma_h^{-1}(\alpha(\sigma))
\end{align}
The exact evaluation of this partition function is already known in literature for algebras of type ABCD and it can be obtained by employing the Weyl character formula and its generalisations. As discussed before, the possible symmetry breaking patterns of the original gauge group for each holonomy configuration fall into an algebra of type ABCD. Therefore, we can get an explicit evaluation of the SCI on each saddle beyond the semiclassical expansion, with the most significant contribution coming from CS partition functions.

The explicit expression of such partition function for each algebra is presented in appendix \ref{AppC}. All of them exhibits similar features. The general structure is

\begin{equation}
\label{eq:Z_ABCD}
    Z_{ABCD(m)_k} = \frac{\exp \left( i \pi f(m,k) \right)}{g(m,k)}\prod_{n \in I_m} 2\sin(\pi \frac{n}{k})^{d(n)},
\end{equation}
where $I_m$ is some subset of consecutive elements of (semi-)integers, typically depending on the rank $m$ of the group, $k$ is the CS level of the theory, while $f(m,k)$ and $g(m,k)$ are two functions depending on the details of $\mathfrak{g}$, with $g(m,k)$ such that $g(m,0) = 0$, while $f(m,k)$ real, so that $\exp(i \pi f(m,k))$ is a phase. The function $d(n)$ represents a possible degeneracy, due to possible multiple occurrences of the same integer n.

A general consequence of (\ref{eq:Z_ABCD}) is that the level plays a crucial role in determining the physical relevance of the saddle point. First, for $k = 0$ the TFT is not well defined. Second, when $k$ lies within $I_m$ the partition function is zero. The only case when (\ref{eq:Z_ABCD}) is non-vanishing is when $k > \mathrm{max}(I_m)$.

Since the CS level is determined by the holonomy configuration of a chosen saddle, we can predict the stability and the contribution of such a saddle to the index only by studying the CS level for the emerging pure CS theories, expanding the effective action for the matrix model near such vacuum.

Focusing first on $\mathcal{N} = 4$ $usp(2N_c)$ SYM, the possible patterns of symmetry breaking found can be divided in two categories $usp(2N_c)\rightarrow usp(2L) \times usp(2K)$ and $usp(2N_c)\rightarrow usp(2L)\times usp(2L)\times su(K) \times u(1)$. For the first case, by inspecting (\ref{eq:Z_usp}) and remembering that the CS levels for the two pure CS theories are defined as in (\ref{eq:levels_usp0half}), we find that $Z^{usp(2(N_c - L))_{k_2}} = 0$ when $0 < L \leq \lfloor N_c/2 \rfloor$. Moreover, under the reflexive symmetry $L \leftrightarrow N_c - L$ the role of $Z^{usp(2L)_{k_1}}$ and $Z^{usp(2(N_c - L)_{k_2})}$ is exchanged and we can conclude that $Z^{usp(2L)_{k_1}} = 0$ when $ \lfloor N_c/2 \rfloor \leq L < N_c$. In addition, it can sporadically happen that the CS level is zero for some saddles with $L \neq 0, 1/2$. Thus, beyond the semiclassical approximation the only non-vanishing saddle arising from the \textcolor{red}{first} family is the one with $N_c$ holonomies at $u_i = 0$ together with its paired one with $N_c$ holonomies at $u_i = 1/2$. All the other saddles give a vanishing partition functions and they are then perturbatively unstable.

The  same argument can be applied to the \textcolor{blue}{second} family of saddles leaving only one non-vanishing saddle with holonomy configuration defined by $N_c$ holonomies at $u_i = 1/4$. 

While the $1/|\tau|^2$ leading order calculation  identifies such two saddles as the dominant contributions to the index, the analysis beyond the leading order shows that they are the only contributions to the index. 

In addition, S-duality cannot hold without the explicit evaluation of the CS partition function obtained by a pertubation close to the saddle. 
This is because S-duality is expected to manifest in the Cardy-like limit as a matching between saddles of the two theories. Then, without an analysis of the subleading contributions in $|\tau|$ of each saddle to the index, not only there is not a clear understanding of the role played by the subleading saddles within the context of S-duality, but even a partial matching between the dominant ones cannot be achieved as discussed in \cite{Amariti:2020jyx}.

The story proceeds in a similar way for the orthogonal $so(2N_c +1)$ case. In this case, we have just one family of saddles with $L$ holonomies at $u_i = 0$ and $N_c - L$ at $u_i = 1/2$, as discussed in Section \ref{sec:SO_saddles}. The original $so(2N_c + 1)$ gauge algebra breaks into $s(o(2L + 1) \times o(2(N_c - L)))$ and a $Z^{s(o(2L + 1)_{k_{1}} \times o(2K)_{k_{2}})}$ factor (with  $k_1$ and $k_2$  defined in (\ref{eq:levels_so0half})) appears in the evaluation of the subleading contributions in $|\tau|$ to the index in the Cardy-like limit.

Using the results presented in appendix \ref{AppC} for the partition function of the CS gauge theories with orthogonal gauge algebra, together with (\ref{eq:levels_so0half}) for the CS levels we  find that the only non-vanishing saddles are the ones with either $N_c$ holonomies at zero or $ L = \lfloor\frac{N_c}{2}\rfloor$ holonomies at $u_i = 0$ and the remaining ones at $u_i = \frac{1}{2}$. These have been already identified as the dominant contributions to the index in the M-wing and W-wing respectively. 

Summarising, S-duality predicts a matching between two pairs of saddles of the two theories, which must hold independently of the regions of charges that we are considering. 
In this sense also the distinction between the W and M shaped regions of the potential is unnecessary, because we have matched the whole expansions in $|\tau|$ in both the wings\footnote{Observe that even if we did not mention the contribution at order $|\tau|$ in our calculation that corresponds, for vanishinbg holonomies, to the supersymmetric Casimir energy \cite{Cassani:2021fyv} and it always matches across dualities. Similarly we have matched that terms across S-duality also in the cases without vanishing holonomies.}.
The SCI for the two distinct 4d S-dual SYM theories reduces to
\begin{itemize}
    \item $usp(2N_c)$: $ \mathcal{I}^{usp(2N_c)} = 2\mathcal{I}_{N_c,0} + \mathcal{I}_{0,0,N_c}$.
    \item $so(2N_c + 1)$: $ \mathcal{I}^{so(2N_c + 1)} = \mathcal{I}_{N_c,0} + \mathcal{I}_{\lfloor \frac{N_c}{2} \rfloor, \lceil \frac{N_c}{2} \rceil}$.
\end{itemize}
  The saddles with vanishing holonomies agree in the two theories, as already discussed in \cite{Amariti:2020jyx}.

For both theories their contribution to the SCI is 
\begin{equation}
   \log \mathcal{I}_{N_c,0} \sim -\frac{i\pi N_c(2N_c+1)}{\tau^2} \prod_{a=1}^3\left(\{\Delta_a\}_\tau - \frac{1+\eta_1}{2}\right) + \log 2 .
    \label{eq:index_final_SO_0}
\end{equation}
This result  holds thanks to the crucial role  played by the evaluation of the CS partition function, responsible in the orthogonal case for the appearance of a  $\log 2$, related to the $\log|G|$ correction to the black hole entropy, discussed in \cite{Amariti:2020jyx} and understood as the presence of a 1-form symmetry.

It remains to show that the saddle with $N_c$ holonomies at $u_i = \frac{1}{4}$ of the $usp(2N_c)$  theory agrees with the corresponding saddle with $ L = \lfloor\frac{N_c}{2}\rfloor$ holonomies at $u_i = 0$ and the remaining ones at $u_i = \frac{1}{2}$ of the $so(2N_c+1)$ theory. 

In the symplectic theory the contribution of the saddle to the index is
\begin{equation}
    \mathcal{I}_{N_c,0} = \tau^N e^{- i \pi \frac{N_c^2}{2}} \mathcal{I}_0 Z_{S^3}^{su(N_c)_{h_1}}Z_{S^3}^{u(1)_{h_2}},
\end{equation}
with 
\begin{equation}
\label{eq:USp_action_1_4}
\begin{split}
\log \mathcal{I}_0 = & -\frac{i\pi N_c(N_c+1)}{4\tau^2}\prod_{a=1}^3\left(\{2\Delta_a\}_\tau - \frac{1+\eta_2}{2}\right) + \frac{i\pi N_c}{\tau^2} \prod_{a=1}^3\left(\{\Delta_a\}_\tau - \frac{1+\eta_1}{2}\right) \\
& + \frac{5i\pi N_c}{12}\left(\eta_1 - (N_c+1)\eta_2 \right) + \frac{i\pi N_c}{2} - N_c\log(\tau),
\end{split}
\end{equation}

For the orthogonal case the general expression (\ref{eq:index_so_final}) reduces to, when $L = \lfloor \frac{N_c}{2} \rfloor$,
\begin{equation}
       \mathcal{I}_{\lfloor \frac{N_c}{2} \rfloor, \lceil \frac{N_c}{2} \rceil } = \tau^{N_c} e^{ - \frac{i \pi N_c^2}{2}} \mathcal{I}_0 Z_{S^3}^{s\left(o(2\lfloor \frac{N_c}{2}\rfloor + 1)_{k_1}\times o(2 \lceil \frac{N_c}{2}\rceil)_{k_2}\right)},
\end{equation}
where 
\begin{equation}
\begin{split}
    \log\mathcal{I}_0 = & \frac{i\pi N_c}{\tau^2} \prod_{a=1}^3\left(\{\Delta_a\}_\tau-\frac{1+\eta_1}{2}\right) - \frac{i\pi N_c(N_c+1)}{4\tau^2} \prod_{a=1}^3\left(\{2\Delta_a\}_\tau-\frac{1+\eta_2}{2}\right)  \\
    &+ \frac{5i\pi N_c}{12}\left(\eta_1 - (N_c+1)\eta_2 \right) + \frac{i\pi N_c}{2} - N_c\log(\tau)   
\end{split}
\end{equation}
exactly matches the same term in the symplectic case.

Assuming S-duality is preserved, then a non-trivial integral identity between products of CS partition functions is expected. Thus, focusing on the regions where $\eta_1 = - \eta_2 =-1$
\footnote{The same identities holds also for 
the case  $\eta_1 = - \eta_2 =1$, that it is related to the one discussed here by a parity transformation.}, it remains to  show that 
\begin{itemize}
\item $N_c=2m$:
\begin{equation}
\label{primaeq}
Z_{S^3}^{s(o(2m+1)_{2m+1}\times o(2m)_{2m+4})}
=
Z_{S^3}^{su(2m)_{2m+4}}Z_{S^3}^{u(1)_{4(2m+1)}}.
\end{equation}
\item $N_c=2m+1$:
\begin{equation}
\label{secondaeq}
Z_{S^3}^{s(o(2m+2)_{2m+2}\times o(2m+1)_{2m+5})}
=
Z_{S^3}^{su(2m+1)_{2m+5}}Z_{S^3}^{ u(1)_{4(2m+2)} }
\end{equation}
\end{itemize}
It turns out that these identities indeed hold. The complete proof is presented in appendix \ref{AppD}.
At last, we achieved a matching between all the saddle points emerging in the Cardy-like limit of the SCI for $\mathcal{N} = 4$ SYM theory with $usp(2N_c)$ and $so(2N_c + 1)$, thus recovering S-duality in the Cardy-like limit of the index for finite $N_c$.

To conclude the analysis we comment on the two special cases of $N_c = 1$ and $N_c=2$, when the algebras isomorphisms between classical Lie algebras extend the matching between the saddle points to all the saddles of the two theories.
We have
\begin{itemize}
    \item When $N_c = 1$, the isomorphism is made explicit upon changing variables in the SCI as $u ^{so} = 2v^{usp}$, implying $\mathcal{I}^{usp(2)} = \mathcal{I}^{so(3)}$. Accounting for the pairing degeneracy of the saddles with $v = 0$ and $v = \frac{1}{2}$, we obtain the expected mapping between saddles:
    \begin{equation}
    \begin{split}
      usp(2) & \hspace{1cm}so(3) \\
         0, \, 1/2 \hspace{0.1cm}  &\longmapsto \hspace{0.3cm} 0 \\
          1/4 \hspace{0.1cm} &\longmapsto \hspace{0.3cm} 1/2.
    \end{split}
    \end{equation}
    \item The case of $N_c = 2$ is physically more interesting, being the only case where a third matching between saddles of the two theories appears.
    
    Again, defining $u_{1,2} ^ {so} = v_1 ^{usp} \pm v_2 ^{usp}$, one can easily show that the two indices (\ref{eq:USp_index}) and (\ref{eq:SO_index}) can be mapped into each others. The corresponding mapping between the saddles in the two theories is the following:
\begin{equation}
\begin{split}
         usp(4) & \hspace{1cm} so(5) \\
         (0, 0), (1/2,1/2) \hspace{0.2cm} &\longmapsto \hspace{0.3cm} (0, 0) \\
         (0, 1/2) \hspace{0.2cm}   &\longmapsto \hspace{0.3cm} (1/2, 1/2)\\
         (1/4, 1/4) \hspace{0.2cm} &\longmapsto \hspace{0.3cm} (0, 1/2).
\end{split}
\end{equation}
Besides the two saddles, already discussed in full generality in the previous section, a third matching appears between the (0,1/2) saddle of $usp(4)$ and the (1/2,1/2) saddle of $so(5)$ as a consequence of the algebra isomorphism relating the two SCIs. However, the matching survives only at order $1/|\tau|^2$  in the Cardy-like expansion as, once the CS partition function contributions are included, an instability emerges in the two saddle points because  the CS levels  (\ref{eq:levels_usp0half}) and (\ref{eq:levels_so0half}) vanish in this case.
\end{itemize}

%%%%%%%%%%%
%%%%%%%%%%%
%%%%%%%%%%%
\section{Conclusions}
%%%%%%%%%%%
%%%%%%%%%%%
%%%%%%%%%%%

In this paper we have studied the fate of S-duality in the Cardy like limit of the SCI of $\mathcal{N}=4$ SYM for the cases with gauge algebra $so(2N_c+1)$ and $usp(2N_c)$.
We have found that such duality is preserved (at finite $N_c$) in a   non-trivial way and only after a complete analysis beyond the leading order $1/|\tau|^2$ in the Cardy-like limit.
The calculation of the subleading corrections in $|\tau|$ requires a saddle point analysis, and,  as we have shown here, there is a lower amount  of saddles in the $so(2N_c+1)$ case
with respect to the ones found in \cite{Amariti:2020jyx} for $usp(2N_c)$. While this is not a problem \emph{per se}, because already at leading level in the W-wing, the index evaluated from two degenerate $usp(2N_c)$ saddles  coincides with the one evaluated on a single $so(2N_c+1)$,
by evaluating  only the leading contribution of each saddle in the Cardy-like limit 
we have not been able to fully match the index of $usp(2N_c)$ with the one of  $so(2N_c+1)$.
Nevertheless we have matched the indices evaluated on the saddles that dominate in the M-wing and the indices evaluated on the saddles that dominate in the W-wing separately.
Even if  the matchings between these saddles holds at finite $N_c$, there can be also other saddles that contribute to the index. We have shown that in general 
these last never contribute to the SCI  because the CS partition function generated from  the expansion in $|\tau|$ vanishes for such saddles. 
We have eventually evaluated the CS partition functions and fully matched the index of the S-dual models in the Cardy-like limit.

Many open questions are leftover.
First  it should be interesting to study the fate of S-duality for models with less supersymmetry and multiple gauge groups. In principle we expect a that the behavior
studied here applies to these cases as well and that similar conclusions can be reached.

Motivated by the study of cases with lower supersymmetry, another analysis that we did not perform here regards the study of the subleading corrections for $\mathcal{N}=4$ $so(2N_c)$ SYM. Even if this is a self dual theory, understanding its behavior may be relevant for extending the analysis to models with $\mathcal{N}=1,2$, where also  $so(2N_c)$ gauge nodes can appear. 

A further generalization regards the fate of Seiberg duality in models with four supercharges. In the toric case one can borrow the results of \cite{GonzalezLezcano:2020yeb}, where the  matching is indeed straightforward in the 
solutions denoted as ``C-center''. Other solutions are nevertheless possible, as discussed in \cite{ArabiArdehali:2019orz}, and it is relevant to understand if they are perturbatively stable, i.e. if they are not vanishing  once the  subleading terms and the CS actions are considered.

Partially related to the last issue, another consequence of our analysis regards the relation between the vacua of the $\mathcal{N}=1^*$ theory on the circle and the vacua extracted from the saddle point analysis of the SCI in the Cardy like limit. We have seen here that such correspondence does not seem to hold in the $usp(2N_c)$ and $so(2N_c+1)$ cases, where the number of solutions does not grow with $N_c$ but it is fixed to $3$ in the first case and $2$ in the second case.
As observed in \cite{Cassani:2021fyv} this value is related to the presence of a 1-form global symmetry, and its value reflects the number of inequivalent lattices of charges of Wilson and 't Hooft lines under the unbroken subgroup of the center of the gauge group. For example in the case of the C-center solutions for $SU(N_c)/\mathbb{Z}_C$ $\mathcal{N}=4$ SYM such number is  $N_c/C$. Furthermore this number corresponds to a logarithmic correction to the contribution of the degenerate saddle to the index. As discussed in \cite{Amariti:2020jyx}, despite the different degeneration  of the saddles, there is a matching of these logs in the index of  $usp(2N_c)$ and   $so(2N_c+1)$ in the W-wing that emerges only after evaluating the CS partition function. In general it would be interesting if  this behavior holds true in general, i.e.if  the number of  lattices associated to the same modding corresponds to the log corrections associated to the index.

To conclude, it is also tempting to associate, along the lines of \cite{Cassani:2021fyv,ArabiArdehali:2021nsx}, the results obtained here to a 3d effective action  emerging from the integration over the massive KK modes coming from the matter multiplets in the reduction on the thermal $S^1$.
While this interpretation is expected for the saddles at vanishing holonomies, it is less clear how to interpret our results for the other saddles along these lines.
Indeed, even in absence of an EFT interpretation, following the discussion in \cite{ArabiArdehali:2021nsx} (see also \cite{Cabo-Bizet:2021plf}), one can associate the saddles at non-vanishing holonomies to the expansion of the index with $e^{2\pi i \tau}$ approaches a root of
unity. In the $su(N_c)$ case  the CS partition function corresponds to an orbifold partition functions on $S^3/\mathbb{Z}_C$. 
Furthermore such solutions are related to the orbifolds of the Euclidean AdS$_5$ BH \cite{Aharony:2021zkr}.
In our case such orbifold interpretation is not straightforward and it deserves further analysis.

%%%%%%%%%%%
\section*{Acknowledgments}
%%%%%%%%%%%%%%%%%%
%
%

The work of A.A., A.Z. . has been supported in part by the Italian Ministero dell'Istruzione, 
Universit\`a e Ricerca (MIUR), in part by Istituto Nazionale di Fisica Nucleare (INFN) through the “Gauge Theories, Strings, Supergravity” (GSS) research project and in part by MIUR-PRIN contract 2017CC72MK-003.

%%%%%%%%%%%
%%%%%%%%%%%
%%%%%%%%%%%
\appendix

\section{The superconformal index}
\label{appA}
In this appendix we survey the main definitions of the superconformal index that we have used in the paper.
The index is defined as
\begin{equation}
\label{eq:index_general}
\mathcal{I} \equiv \Tr(-1)^F e^{-\beta  H} p^{J_1+\frac{R}{2}} q^{J_2 + \frac{R}{2}}
\prod_{f=1}^{\text{rk} \mathbf{F}} v_f^{q_f} 
\end{equation}
In this trace formula we $J_i$ are the angular momenta on the $S^3$, $R$ is the R-charge and $q_f$
are the flavor charges of the rank $\mathbf{F}$ flavor symmetry group $\mathfrak{f}$.
The fugacities of these symmetries are denoted as  $p, q$ and $v_f$ respectively.
Instead of the trace formula it is useful to define the index for a gauge theory in terms of a matrix integral over the 
holonomies of the gauge algebra:
\begin{equation}
\label{inde}
\mathcal{I} = \frac{(p;p)_\infty^{\text{rk} \mathfrak{g}} (q;q)_\infty^{\text{rk} \mathfrak{g}}}{|W_{\mathfrak{g}}|}
\oint_{T^{\text{rk}  \mathfrak{g}}}  \prod_{i=1}^{\text{rk}  \mathfrak{g}} \frac{dz_i}{2 \pi i z_i} 
\frac
{\prod_{a=1}^{n_\chi}
\prod_{\rho_a} 
\Gamma_e((pq)^{{R_a}/{2}} z^{\rho_{\mathfrak{g}}^{a}} v^{\rho_{\mathfrak{f}}^{a}})}
{\prod_\alpha \Gamma_e(z^{\alpha_{\mathfrak{g}}})}
\end{equation}
where $\rho_{\mathfrak{g,f}}^a$ represent the weights for the chiral multiplet  gauge and the flavor.

In the Cardy-like limit it is more convenient to work explicitly with the chemical potentials conjugated to the charges of the theory. Therefore we define
\begin{equation}
    p \equiv e^{2\pi i \tau}, \hspace{0.5cm} q \equiv e^{2\pi i \sigma}, \hspace{0.5cm} v_j \equiv e^{2\pi i \xi_j}, \hspace{0.5cm} z_i \equiv e^{2\pi i u_i}.
\end{equation}
From \ref{eq:index_general} we can read off the chemical potential for the R-charge 
\begin{equation}
    \nu_R \equiv \frac{1}{2}(\tau + \sigma).
\end{equation}
In the literature it is pretty common to encode all the charges associated with the global symmetries of the theory in a new set of fugacities $y_a$ together with the charges $\Delta_a$ defined by
\begin{equation}
    y_a \equiv e^{2\pi i \Delta_a}  \equiv ((pq)^{{R_a}/{2}}  v^{\rho_{\mathfrak{f}}^{a}}),
\end{equation}
The charges $\Delta_a = \rho_f^a(\mathbf{\xi}) + \nu_R R_a$ encode all the information about the flavour and R charges of the theory.

\section{Asymptotic formulas}
\label{appB}

In this appendix we collect the main formulas for the hypergeometric functions and their asymptotic expansions needed to perform a saddle-point evaluation of the SCI in the Cardy-like limit.

Let $\tau, \sigma \in \mathbb{H}$ and $p = e ^ {2\pi i \tau}$, $ q = e ^ {2\pi i \sigma}$.
The elliptic gamma function is defined as the infinite product 
\begin{equation}\label{GammaE}
    \Gamma_e(z;p,q) \equiv \Gamma_e(z) \coloneqq \prod_{i,j = 0}^{\infty} \frac{1 - p^{j+1}q^{k+1}/z}{1 - zp^j q ^k}.
\end{equation} 

We also define the modified elliptic gamma function as
\begin{equation}
    \Tilde{\Gamma}_e(u; \tau, \sigma) \equiv \Tilde{\Gamma}_e(u) \coloneqq \Gamma_e (e^{2\pi i u};e^{2 \pi i \tau}, e^{2 \pi i \sigma}).
\end{equation}

The Pochhammer symbol is defined for complex $z,q$ with $|q| < 1$ by
\begin{equation}
\label{Poch}
    (z;q)_\infty \coloneqq \prod_{j = 0} ^ {\infty} \left( 1 - z q^k \right).
\end{equation}

We can then define the elliptic function $\theta_0$ 
\begin{equation}
    \theta_0(u;\tau,\sigma) = (u;q)_\infty(q/u;q)_\infty = \prod_{ j = 0 } ^{\infty} \left( 1 - e ^ {2\pi i ((j + 1) \tau - u)}\right) \left( 1 - e^{2\pi i (u + j \tau) }\right)
\end{equation}

and for our purposes it is enough to remember that it satisfies 
\begin{equation}
\label{eq:gamma_theta_rel}
\log \theta_0 (u) = -\log \tilde{\Gamma}_e(u).
\end{equation}
In the Cardy-like limit the asymptotic behaviour of these functions can be written introducing the $\tau-$modded value of a complex number:
\begin{equation}
    \{u\}_\tau \equiv u - \lfloor \Re (u) - \cot(\arg(\tau))\Im (u) \rfloor,
\end{equation}

For $u \in \mathbb{R}$ it reduces to the ordinary fractional part $u - \lfloor u \rfloor$.

Writing $u \in \mathbb{C}$ as $u = \Tilde{u} +\tau\check{u}$ with $\Tilde{z},\check{z}\in\mathbb{R}$, the $\tau-$modded value satisfies
\begin{equation}
    \{u\}_\tau = \{\Tilde{u}\} + \tau\check{u},
\end{equation}

Moreover from the definition it follows that 

\begin{equation}
    \{-u\}_\tau=\begin{cases}
    1-\{u\}_\tau \hspace{0.5cm}\Tilde{u}\not\in\mathbb{Z}\\
    -\{u\}_\tau \hspace{0.9cm}\Tilde{u}\in\mathbb{Z}.
    \end{cases}
\end{equation}

Then, as $|\tau|\rightarrow 0$ with $\Im{\tau} > 0$ fixed, we have the following asymptotic behaviours

\begin{equation}
    \log \left(q ; q \right)_\infty \sim -\frac{i \pi}{12} \left( \frac{1}{\tau} + \tau \right) - \frac{1}{2} \log( -i \tau) + \mathcal{O} \left( e ^ {\frac{ 2 \pi \sin{\arg(\tau)} }{|\tau|}}\right),
    \label{eq:asymptotic_Poch}
\end{equation}

\begin{equation}
  \begin{split}
    \log \theta_0 (u;\tau) \sim & \frac{i \pi }{\tau} \{u\}_\tau (1-\{u\}_\tau) + i \pi \{u\}_\tau-\frac{i \pi }{6\tau} \left(1 + 3\tau + \tau^2\right) + \\
    & + \log \left[ \left(1 - e ^ { - \frac{ 2\pi i }{\tau} \{u\}_\tau} \right) \left(1 - e ^{ - \frac{2\pi i}{\tau}(1 - \{u\}_\tau)} \right) \right] + \mathcal{O} \left(e^{ \frac{2 \pi \sin{ \arg(\tau)} }{|\tau|}} \right),
  \end{split}
    \label{eq:asymptotic_theta}
\end{equation}

\begin{equation}
    \log\Tilde{\Gamma}(u;\tau)\sim 2\pi i Q(\{u\}_\tau;\tau) + \mathcal{O} \left( |\tau|^{-1} e^{ \frac{ 2 \pi \sin{\arg(\tau)} }{|\tau|} \min ( \{u\}_\tau, 1 - \{u\}_\tau)} \right),
    \label{eq:asymptotic_gamma}
\end{equation}
provided that $\Tilde{u}\not\in \mathbb{Z}$, with $Q(u)$ defined as
\begin{equation}
    Q(u;\tau) = -\frac{B_3(u)}{6\tau^2} + \frac{B_2(u)}{2\tau} - \frac{5B_1(u)}{12} + \frac{\tau}{12},
\end{equation}

where $B_n(u)$ are the Bernoulli polynomials
\begin{equation}
    B_1(u) = u - \frac{1}{2}, \hspace{1cm} B_2(u) = u^2 - u + \frac{1}{6}, \hspace{1cm} B_3(u) = u^3 - \frac{3}{2}u^2 + \frac{u}{2}.
\end{equation}
The Bernoulli polynomials (and their modded version $B(\{x\}_\tau)$) satisfy the following identity, known as Raabe's formula

\begin{equation}
    \sum_{J=0}^{C-1}B_n\left(\frac{J}{C} + u\right) = \frac{1}{C^{n-1}}B_n\left(Cu\right),
\end{equation}
through which we expressed the effective actions in terms of products of $\{C\Delta\}_\tau$ terms, with $C = 1, 2, 4$.

\section{$Z_{S^3}$ for pure 3d $\mathcal{N}=2$ CS theories with ABCD gauge algebra}
\label{AppC}
%%%%%%%%%%%
%%%%%%%%%%%
%%%%%%%%%%%

In this appendix we collect some useful formulas on the exact evaluation of the 3d partition function 
of the three sphere partition function 3d of pure CS gauge theories with gauge algebra $\mathfrak{g}$ of ABCD type.
The integral formula corresponds to a matrix integral of the form
\begin{align}
\label{genZ}
Z_{\mathfrak{g}} = \frac{1}{|W| } \int \prod_{i=1}^{\text{rk}\mathfrak{g}}
d \sigma_i
e^{\frac{i \pi k \sigma_i^2 }{\omega_1 \omega_2} }
\prod_{\alpha(\sigma)} \Gamma_h^{-1}(\alpha(\sigma))
\end{align}
where $\alpha(\sigma)$ represent the simple roots of the algebra and $\Gamma_h$ are 
hyperbolic gamma functions 
\begin{equation}\label{hyp}
\Gamma_h(z) \equiv \prod_{m,n=1}^\infty \frac{(n+1)\omega_1+(m+1)\omega_2 - z}{n\omega_1 + m\omega_2}
\end{equation}
Expression (\ref{genZ}) can be rewritten in terms of hyperbolic sines, as in the main text, by employing the following property of the hyperbolic gamma functions
\begin{equation}
\label{hyp_sin}
    \frac{1}{\Gamma_h(x)\Gamma_h(-x)} = - 4 \sin(\frac{\pi x}{\omega_1})\sin(\frac{\pi x}{\omega_2}).
\end{equation}
We have observed in the body of the paper that the Cardy-like limit of the SCI gives rise 
(for charges $\Delta_a \neq 0,2$) to a matrix integral of the type (\ref{genZ}) for 
pure CS gauge algebras of $\mathfrak{g}$ of ABCD type.

Such partition functions can be exactly evaluated and the results have already been obtained in the literature. Here we collect these results, because the exact evaluations  of $Z_{\mathfrak{g}} $  has allowed us to perform the precision checks on S-duality. Furthermore we restrict to the case of $S^3$, setting $\omega_1 = \omega_2$ because we 
have studied the case with collinear angular momenta in the body of the paper.

Let us start surveying the various results. The evaluation of the partition function for pure CS $\mathfrak{g}=u(N_c)$ at level $k$ was performed in \cite{Kapustin:2009kz}. The final formula is
\begin{align}
Z_{u(N_c)_k} =&\
\frac{e^{\frac{i \pi  N_c \left(3 k N_c-2 N_c^2-6 k+2\right)}{12 k}}}{k^{\frac{N_c}{2}}}
\prod _{m=1}^{N_c-1} \left(2 \sin \left(\frac{\pi  m}{k}\right)\right)^{N_c-m}
\end{align}
It is also possible to relate the $u(N_c)$ case to the $su(N_c)$ one, thanks to the formula
\begin{align}
Z_{su(N_c)_k}=&\
\sqrt{\frac{k}{i N_c}}
Z_{u(N_c)_k}
\end{align}
Such distinction is important in our analysis, because we often deal with $su(N_c) \times U(1)$ gauge theories, with different CS level for the abelian factor.

The partition function for 3d pure CS on $S^3$ with $\mathfrak{g} = usp(2N_c)$ at level $k$ is \cite{VanDeBult}
\begin{align}
\label{eq:Z_usp}
Z^{usp(2N_c)_k} = &\
\frac{\exp \left(- \frac{ i \pi N_c (2 + 6 N_c + 4 N_c^2 + 6 k + 3 |k|)} {12 k}\right)}{(2 |k|)^\frac{N_c}{2}}  \cdot \nonumber \\
& \cdot \!\!\! \! \prod_{1\leq j<\ell \leq N_c} 4 \sin \left( \frac{\pi(j + \ell)}{2k} \right) \sin \left(\frac{\pi(j - \ell)}{2k} \right) 
\prod_{j=1}^{N_c} 2 \sin \left(\frac{ \pi j}{k} \right) \quad . 
\end{align}

To conclude.the survey we consider the orthogonal cases, studied in \cite{Amariti:2020jyx}.
The case of $\mathfrak{g}=so(2N_c+1)$ at level $k$ gives 
\begin{align}
\label{eq:Z_soOdd}
Z^{so(2N_c+1)_k} =&\
\frac{\exp \left(\frac{i \pi  N_c \left(12 k N_c-8 N_c^2-9 |k| +2\right)}{12 k}\right)}{| k| ^{\frac{N_c}{2}}}  \cdot \nonumber \\
& \cdot \!\!\! \! \prod_{1\leq j<\ell \leq N_c} 4 \sin \left( \frac{\pi(j+\ell-1)}{k} \right) \sin \left(\frac{\pi(j-\ell)}{k} \right) 
\prod_{j=1}^{N_c} 2 \sin \left(\frac{ \pi (2j-1)}{2k} \right)\, 
\end{align}
while for  of $\mathfrak{g}=so(2N_c)$ at level $k$ we have
\begin{align}
\label{eq:Z_soEven}
Z^{so(2N_c)_{k}}_{S_b^3} =&\ 
\frac{\exp \left(\frac{i \pi  N_c \left(12 k N_c-8 N_c^2-9 |k|+2\right)}{12 k}\right)}{|k|^\frac{N_c}{2}}
\cdot
\nonumber \\ 
&\cdot
\prod_{1 \leq j <\ell <N_c} 4 \sin \left( \frac{\pi(j+\ell-2}{|k|} \right) \sin \left( \frac{\pi(j-\ell)}{|k|} \right) \ .
\end{align}

%%%%%%%%%%%%%%%%%%%%%%%%%%%%%%%%%%
%%%%%%%%%%%%%%%%%%%%%%%%%%%%%%%%%%
\section{Proof of (\ref{primaeq})}
\label{AppD}
%%%%%%%%%%%%%%%%%%%%%%%%%%%%%%%%%%
%%%%%%%%%%%%%%%%%%%%%%%%%%%%%%%%%%
In this appendix we prove the relation (\ref{primaeq}) between the partition functions of pure CS gauge theories that 
we have used in the body of the paper in order to show how  S-duality is preserved in the Cardy-like limit in the region where the index is dominated by the $W$-wing shaped potential.
As discussed in the paper we have identified two different possibilities for the $usp(2N_c)/so(2N_c+1)$ duality, depending on the parity of $N_c$.
For $N_c=2m$ the expected relation is (\ref{primaeq}) that we reproduce here for the ease of the reader
\begin{equation}
\label{USp4m}
Z_{S(O(2m+1)_{2m+1}\times O(2m)_{2m+4})}
=
Z_{U(2m)_{2m+4,4(2m+1)} }
\end{equation}
while for $N_c=2m+1$ the expected relation is (\ref{secondaeq})
\begin{equation}
\label{USp4m2}
Z_{S(O(2m+2)_{2m+2}\times O(2m+1)_{2m+5})}
=
Z_{U(2m+1)_{2m+5,4(2m+2)} }
\end{equation}
Two comments are in order.
First the normalization of the $U(1)$ factor follows the conventions of appendix A of \cite{Amariti:2020xqm}. Second, the partition function for the orthogonal case has been denoted here as schematically as $S(O(n) \times O(m))$ but it coincides with the one of $so(n) \times O(m)$ and $O(n) \times so(m)$.

The two identities   (\ref{USp4m}). and  (\ref{USp4m2}) can be shown explicitly.
In the following  we give a direct derivation of  (\ref{USp4m}).
An analogous derivation holds for (\ref{USp4m2}).

We start observing that the partition function   $Z_{so(2m+1)_{2m+1}}$ 
can be evaluated, inferring the results from  the evaluation  of
$Z_{so(2n+1)_{2n-1}}$ given in \cite{Amariti:2020jyx}.
We have
\begin{equation}
\label{SOnostra}
Z_{so(2m+1)_{2m+1}}=
\frac{e^{\frac{1}{12} i \pi  m (2 m-1)}}{\sqrt{2 m+1}}
\end{equation}

On the other hand we can estimate the relation between the products of trigonometric functions 
that are inside the $so(2m)$ and the $U(2m)$ partition functions.
They are
\begin{equation}
\label{eqSO}
\mathcal{P}_{SO}  \equiv \prod _{p=1}^m \prod _{q=p+1}^m 4 \sin \left(\frac{\pi  (p-q)}{2 (m+2)}\right) \sin \left(\frac{\pi  (p+q-2)}{2 (m+2)}\right)
\end{equation}
and
\begin{equation}
\label{eqU}
\mathcal{P}_{SU}  \equiv \prod _{p=1}^{2m-1} \left(2 \sin \left(\frac{\pi  p}{2m+4}\right)\right)^{2m-p}=\prod _{p=1}^{2m} \prod _{q=p+1}^{2m} 2 \sin \left(\frac{\pi  (p-q)}{m+4}\right)
\end{equation}
respectively

The ratio between such quantities can be simplified by using the partition functions of the pure 3d $\mathcal{N}=2$ CS $U(n)_{n}$ and $so(2n)_{2(n-1)}$ theories. 
They first one can be read from   \cite{Amariti:2020jyx} and  for   $Z_{so(2m+6)_{2(m+2)}}$ it gives 
\begin{equation}
\label{noiSOe}
\prod _{p=1}^{m+3} \prod _{q=p+1}^{m+3} 4 \sin \left(\frac{\pi  (p-q)}{2 (m+2)}\right) \sin \left(\frac{\pi  (p+q-2)}{2 (m+2)}\right)
=
2 e^{\frac{1}{2} i \pi  \left(m^2+m+2\right)} (2 m+4)^{\frac{m+3}{2}}
\end{equation}
while the second one, for $Z_{U(m+4)_{m+4}}$, can be read from \cite{Kapustin:2009kz} and it gives
\begin{equation}
\label{Kap}
\prod _{p=1}^{2m+4} \prod _{q=p+1}^{2m+4} 2 \sin \left(\frac{\pi  (p-q)}{2m+4}\right)
=
(2m+4)^{m+2}
\end{equation}
Using (\ref{noiSOe}) we simplify (\ref{eqSO}) as 
\begin{equation}
\label{eqSO2}
\mathcal{P}_{SO}  = 
2 e^{\frac{1}{2} i \pi  \left(m^2+m+2\right)} (2 m+4)^{\frac{m+3}{2}} \Theta_{SO}
\end{equation}
with
\begin{eqnarray}
\Theta_{SO}&=&
\frac{1}{\prod _{p=1}^m \prod _{q=m+1}^{m+3} 4 \sin \left(\frac{\pi  (p-q)}{2 (m+2)}\right) \sin \left(\frac{\pi  (p+q-2)}{2 (m+2)}\right)} 
\nonumber \\
&\times & 
\frac{1}{\prod _{p=m+1}^{m+2} \prod _{q=p}^{m+3} 4 \sin \left(\frac{\pi  (p-q)}{2 (m+2)}\right) \sin \left(\frac{\pi  (p+q-2)}{2 (m+2)}\right)}
\end{eqnarray}
Using (\ref{Kap}) we simplify  (\ref{eqU}) as 
\begin{equation}
\label{eqU2} 
\mathcal{P}_{SU}  = 
(2m+4)^{m+2} \Theta_{SU}
\end{equation}
with
\begin{equation}
\Theta_{SU}=\frac{1}
{\prod _{p=1}^{2m} \prod _{q=2m+1}^{2m+4} 2 \sin \left(\frac{\pi  (p-q)}{2m+4}\right) \cdot \prod _{p=2m+1}^{2m+3} \prod _{q=p+1}^{2m+4} 2 \sin \left(\frac{\pi  (p-q)}{2m+4}\right)}
\end{equation}
Next we want to show that
\begin{equation}
\label{finaltheta}
\frac{\Theta_{SU}}{\Theta_{SO}} =\frac{(-1)^{m+1}}{m+2 }
\end{equation}
In order to evaluate this ratio we start observing that
\begin{equation}
 \frac{\prod _{p=m+1}^{m+2} \prod _{q=p}^{m+3} 4 \sin \left(\frac{\pi  (p-q)}{2 (m+2)}\right) \sin \left(\frac{\pi  (p+q-2)}{2 (m+2)}\right)
 }{
  \prod _{p=2m+1}^{2m+3} \prod _{q=p+1}^{2m+4} 2 \sin \left(\frac{\pi  (p-q)}{2m+4}\right)}=-1
\end{equation}
We are then left then with
\begin{eqnarray}
\label{finaltheta2}
\frac{\Theta_{SU}}{\Theta_{SO}} &= &- \frac{\prod _{p=1}^m \prod _{q=m+1}^{m+3} 4 \sin \left(\frac{\pi  (p-q)}{2 (m+2)}\right) \sin \left(\frac{\pi  (p+q-2)}{2 (m+2)}\right) }{\prod _{p=1}^{2m} \prod _{q=2m+1}^{2m+4} 2 \sin \left(\frac{\pi  (p-q)}{2m+4}\right) }
\nonumber \\
&=&
\frac{(-1)^{m+1} }
{\sin \left(\frac{4 \pi }{2 m+4}\right) \cdot \prod _{p=1}^m 2 \sin \left(\frac{\pi  (p+1)}{2 m+4}\right) \cdot \prod _{p=1}^{m-1} 2 \sin \left(\frac{\pi  p}{2 m+4}\right)}
\end{eqnarray}
In order to conclude the proof of (\ref{finaltheta}) we need to estimate the denominator of (\ref{finaltheta2}).
This can be done by using the relations
\begin{equation}
\sin \left(\frac{4 \pi }{2 m+4}\right)=4 \sin \left(\frac{\pi  (2 m+3)}{2 (m+2)}\right) \sin \left(\frac{\pi  (m+3)}{2 (m+2)}\right) \sin \left(\frac{\pi  (m+4)}{2 (m+2)}\right)
\end{equation}
and
\begin{equation}
\prod _{p=1}^m  \sin \left(\frac{\pi  (p+1)}{2 (m+2)}\right)=\prod _{p=m+3}^{2 m+2} \sin \left(\frac{\pi  p}{2 (m+2)}\right)
\end{equation}
Such that the denominator of (\ref{finaltheta2}) becomes
\begin{equation}
\frac{1}{2} \prod _{p=1}^{2 m+3} 2 \sin \left(\frac{\pi  p}{2 m+4}\right) = m+2
\end{equation}
Then by plugging this result in  (\ref{finaltheta2})  we arrive at  (\ref{finaltheta}).
Eventually  we plug (\ref{SOnostra}) and  (\ref{finaltheta}) in  (\ref{USp4m}) such to verify the latter.

To conclude we have not presented the explicit derivation of 
(\ref{USp4m2}), because it can be derived along the same lines of the analysis performed in this appendix.

\bibliographystyle{JHEP}
\bibliography{biblio.bib}
\end{document}